\documentclass[runningheads]{llncs}
\usepackage[T1]{fontenc}
\usepackage{hyperref}
\usepackage{caption}
\usepackage{subcaption}
\usepackage{algorithm}
\usepackage{algcompatible}
\usepackage{algorithmicx}
\usepackage{algpseudocode}
\usepackage{comment}
\usepackage{enumitem}
\usepackage{graphicx}%
\usepackage{multirow}%
\usepackage{amsmath,amssymb,amsfonts}%
\usepackage{mathrsfs}%
\usepackage{xcolor}%
\usepackage{textcomp}%
\usepackage{manyfoot}%
\usepackage{booktabs}%
\usepackage{listings}%
\usepackage{hhline}
\begin{document}
\title{Committee Configuration Optimization for Parallel Byzantine Consensus in a Trusted Execution Environment}
\titlerunning{Committee Configuration Optimization for Parallel BFT in TEE}
%
\author{Yifei Xie\inst{1} \and
Btissam Er-Rahmadi\inst{2} \and
Xiao Chen\inst{3} \and
Tiejun Ma\inst{1} \and
Jane Hillston\inst{1}}
\authorrunning{Xie et al.}
%
\institute{School of Informatics, University of Edinburgh, UK \\
\email{yifei.xie@ed.ac.uk}\and
Independent Researcher, Edinburgh, UK \and
School of Computing and Mathematical Sciences, University of Leicester, UK}
\maketitle              
\begin{abstract}
Parallel Byzantine Fault Tolerant (BFT) protocols based on committee-based sharding improve scalability but weaken safety since smaller node groups are responsible for consensus. Recent approaches integrate trusted execution environments (TEEs) into parallel BFT frameworks to enhance safety. While the scalability and safety issues are addressed by trusted parallel BFT, existing committee configuration methods often rely on randomized assignment, which can degrade performance.
This paper proposes a committee configuration optimization (CCO) model based on mixed integer programming to improve transaction performance for trusted parallel BFT\@. The model considers communication delays and node failure rates to determine an optimal committee configuration that minimizes transaction latency under both normal operations and scenarios of trusted hardware failures. We integrate CCO into a trusted parallel BFT protocol and evaluate the performance on Microsoft virtual machines. Experimental results demonstrate 15\% and 21\% improved transaction throughput under normal operations and fallback process, respectively, highlighting the benefits of optimization-driven committee configuration in trusted parallel BFT systems.

\end{abstract}
\section{Introduction} \label{sec:intro}
Partitioning Byzantine Fault Tolerant (BFT) systems into multiple parallel committees is considered a promising strategy \cite{kokoris2018omniledger,chen2023parbft,zhang2025scaling} for overcoming the scalability challenges of traditional BFT protocols. Classical BFT protocols suffer from $\mathcal{O}(N^2)$ communication complexity with $N$ participating nodes as they typically require $3f+1$ nods to tolerate $f$ malicious nodes \cite{lamport2019byzantine}. While sharding is introduced to mitigate scalability issues, it often relies on smaller groups of nodes (e.g., site groups in \cite{amir2008steward} or upper-layer/global committees in \cite{luu2016secure}) to ensure transaction total order.  This can compromise system safety compared to the classic BFT protocols like PBFT \cite{castro1999practical}, where all nodes are responsible for ensuring total order safety. Driven by the growing adoption of BFT as a consensus protocol in blockchain systems \cite{xu2024x,jia2024estuary}, the requirements for BFT algorithms have increasingly emphasized higher performance and safety \cite{androulaki2018hyperledger}. Consequently, hardware assisted security solutions, particularly trusted execution environments (TEEs), have been integrated into BFT and parallel BFT to enhance safety \cite{veronese2011efficient,liu2018scalable,chen2024parallel}. The TEE/hardware-based solution can guarantee consensus safety based on a constrictive lower bound (i.e., $2f +1$) to tolerate $f$ faulty servers.

Despite the safety and scalability gains provided by trusted parallel BFT, these systems still face significant performance limitations due to inefficient committee configuration methods.  
Most current BFT protocols use random configuration methods to secure committee membership  \cite{micali1999verifiable}, which reduces consensus responsiveness. In real-world environments, responsiveness is heavily influenced by node-to-node message delays and node reliability. For instance, when slow nodes are spread across committees, it can delay consensus in each committee and the overall process, increasing latency. This issue is exacerbated in large-scale peer-to-peer (P2P) systems, where network conditions vary and node availability fluctuates.
Furthermore, the introduction of TEEs brings a configuration challenge of hardware failure. If the trusted status of a TEE is compromised or the hardware itself fails, the committee's resilience bound is invalidated. This necessitates a \emph{fallback} process where the system must reconfigure to $3f+1$ mode to maintain safety. Current configuration methods largely overlook this scenario, failing to account for the performance degradation that occurs when trusted hardware becomes unavailable or requires adaptive reconfiguration. Consequently, an optimized approach to committee configuration is essential for maximizing performance in a trusted parallel BFT system.

Recent efforts to optimize committee configurations have explored several methodologies. Previous research utilizes machine or reinforcement learning to tune parameters such as block size and shard counts \cite{yun2020dqn,zhang2025scaling}, while other work, such as GeoSharding \cite{ruparel2020geosharding}, focuses on geographical mapping and leader election. 
Alternative approaches such as the score-based election in \cite{liu2025abse}, propose an adaptive baseline score-based election approach for leader configuration in BFT systems.
More recently, mathematical programming has been applied to optimize parallel BFT protocol performance in various scenarios \cite{chen2023parbft,xie2022stochastic,xie2025stochastic}. Chen et al. \cite{chen2023parbft} formulate a bi-level mixed-integer linear programming (BL-MILP) model to determine the optimal committee configuration that maximizes operation performance. Xie et al. \cite{xie2025stochastic} take the uncertainty of network conditions into account and proposed a stochastic programming (SP) model to capture random communication delays and failure rate to optimally configure the committee membership of parallel BFT\@. In addition, a reputation evaluation approach proposed in \cite{yang2025gaussian}, integrates Gaussian reputation score evaluation with a proof-of-stake (PoS) mechanism to reconfigure candidate members for committee-based BFT. Xie et al. \cite{xie2026optimizing} first consider optimizing the view change process by efficiently configuring the committee leaders in parallel consensus. While these studies have considered optimizing BFT configuration, the specific challenge of configuring committee membership for TEE-based parallel BFT remains underexplored.

In order to improve the performance of parallel BFT in a trusted environment, we propose a committee configuration optimization (CCO) model based on mixed integer programming (MIP). The CCO model considers communication performance efficiency and fault tolerance to determine the optimal configuration across multiple committees. By accounting for node-to-node delay and failure rate, the proposed model minimizes transaction operation latency under both normal operations and worst scenario when trusted hardware fails. The main contribution of this work is summarized as follows:
\begin{itemize}
\item Our proposed model is the first to address performance issues in a trusted parallel BFT environment, optimizing committee membership configuration for parallel BFT in TEEs.  
\item The proposed model handles adaptive optimization for the fallback process, dynamically reconfiguring committees to maintain system reliability and performance when TEE nodes fail in their respective committees.  
\item We integrate our CCO model into a trusted parallel BFT protocol (TopBFT proposed by \cite{chen2024parallel}) and perform experiments on Microsoft virtual machines. Real-world experiment results demonstrate that CCO model improves performance under various conditions. These results show the effectiveness of the CCO model in configuring parallel BFT in trusted environments.  
\end{itemize}

\section{Related Work} \label{sec:ro-lr}
To address scalability and performance issues, researchers have explored solutions that involve deploying multiple consensus committees. This approach splits the processing of transactions among smaller groups of nodes, which can operate on consensus in parallel.
Notable examples of such solutions include ELASTICO \cite{luu2016secure}, which employs a hierarchical structure supported by a top-layer committee and several sub-committees, and Omniledger \cite{kokoris2018omniledger}, which utilizes bias-resistant public randomness to generate committees. Additionally, Rapidchain \cite{zamani2018rapidchain} is the first sharding-based public blockchain protocol primarily focused on developing an optimal intra-committee consensus algorithm. ParBFT \cite{chen2023parbft} integrates a sharding scheme with a multi-signature technique to partition the consensus network into several committees, allowing transactions to be executed in parallel. 

TEEs such as Intel's SGX \cite{cryptoeprint:2016/086} provide protected memory and isolated execution so that the regular operating system or applications can neither control nor observe the data being stored or processed inside them. TEEs also allow remote verifiers to ascertain the current configuration and behavior of a device via \emph{remote attestation}. In other words, TEE can only crash but not be Byzantine.
Previous work showed how to use hardware security to reduce the number of replicas and/or communication phases for BFT protocols \cite{veronese2011efficient,kapitza2012cheapbft,liu2018scalable}. 
MinBFT \cite{veronese2011efficient} improves PBFT using a \emph{trusted counter service} to prevent equivocation by faulty replicas.  
By binding each consensus message to a unique, monotonic counter maintained within the TEE, replicas cannot assign the same counter value to different messages. This reduces the required replicas from $3f+1$ to $2f+1$.
CheapBFT \cite{kapitza2012cheapbft} uses TEEs in an optimistic BFT protocol that only requires $f+1$ active replicas to execute in the absence of faults, while other $f$ passive replicas just maintain state updates provided by the active replicas. 
FastBFT \cite{liu2018scalable} is another hardware secured BFT algorithm, which uses TEEs to secure counters and a secret-sharing scheme to achieve BFT consensus with $2f + 1$ replicas. 
Chen et al. \cite{chen2024parallel} propose TopBFT by integrating parallel BFT in TEEs, thereby enhancing both the safety and scalability.


\section{Preliminaries} \label{sec:overview}


\subsection{Consensus Parallelism in TEE}
\begin{figure}[!t]
	\centering
	\includegraphics[scale=0.24]{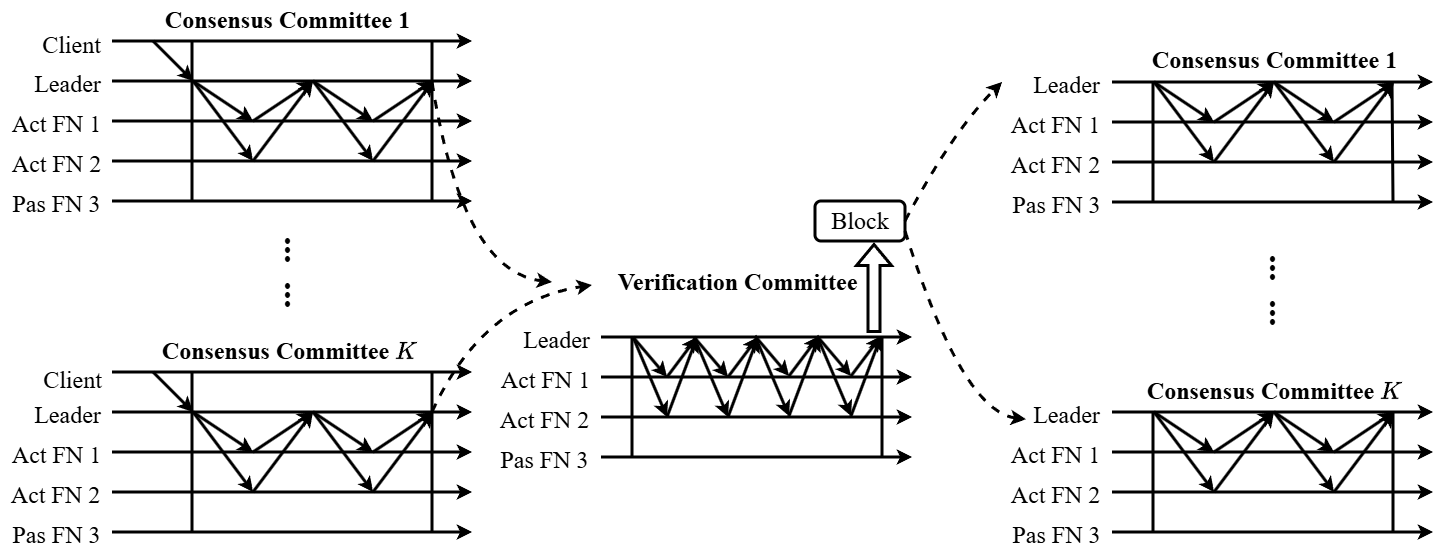}
	\caption{Normal case operations of TopBFT.}
	\label{fig:consensus-phase}
    \vspace{-3mm}
\end{figure}

TopBFT, as depicted in Fig.~\ref{fig:consensus-phase}, implements parallel Byzantine consensus through several committees which can be categorized into two types: $\mathcal{S}_v$, a \emph{verification committee} responsible for committee configuration, secret management, and total order consensus; and $\mathcal{S}_c$, \emph{consensus committees} responsible for local-order consensus and total order verification. 
The clients connect with consensus nodes via communication networks.

TopBFT supports concurrent local consensus in $\mathcal{S}_c^k$ and batches processing for all local transactions in the $\mathcal{S}_v$. We assume that there are $K + 1$ committees, i.e., one $\mathcal{S}_v$ and $K$ $\mathcal{S}_c^k$s, i.e. $1 \!\leq \!k \!\leq \!K$. Each $\mathcal{S}^k_c$ contains a leader node and several follower nodes. The $\mathcal{S}_c^k$ accepts a request from the client, starts local consensus, and orders the request by assigning a unique sequence in the \emph{Pre-prepare} phase, then multicasts the ordered request to other follower nodes for the following operations. Hence, the parallel consensus allows each $\mathcal{S}_c^k$ to accept requests and run consensus in parallel, which significantly scales Byzantine consensus in a large-scale node set. As illustrated in Fig.~\ref{fig:consensus-phase}, during normal operation, each committee reaches consensus using only $2f$ active TEE-enabled follower nodes. In the event of a TEE failure, the $f$ passive nodes are activated, transitioning the committee to a $3f+1$ configuration to maintain system safety.

\subsection{Algorithm Operations}
If we view the system from a single transaction point of view, better performance means that the time spent to commit a transaction successfully should be as small as possible. The total time spent on a successful transaction depends on several phases of the consensus protocol. A successful round of the TopBFT consensus protocol works as follows:
\begin{itemize}
    \item[1)] \textit{Request}: A client sends a request to invoke a service transaction to the leader node of a consensus committee $\mathcal{S}_c^k$.
    \item[2)] \textit{Pre-prepare}: The leader node of the $\mathcal{S}_c^k$ that received the client's request multicasts the request message to all the follower nodes in its committee.
    \item[3)] \textit{Prepare}: All the followers execute the request and send their signed messages to the leader node. The latter aggregates all these messages into a multi-signature and sends it to the verification committee.
    \item[4)] \textit{Verifying and Total Ordering}: The verification committee collects messages from consensus committees, then verifies these messages and builds a block by totally ordering all the transactions from consensus committees.
    \item[5)] \textit{Commit}: Once the multi-signature is verified, the verification committee sends a commit message to the $\mathcal{S}_c^k$ to notify it that at least two-thirds of nodes in the $\mathcal{S}_c^k$ operate correctly.
    \item[6)] \textit{Reply}: The leader node of $\mathcal{S}_c^k$ verifies the multi-signature and sends a reply to the client, thus completing this consensus round.
\end{itemize}

Times spent on phases 1 and 6 are exogenous to the network configuration, since they are determined by the network delay between the client and the consensus committee. Therefore, we exclude these two phases when we formulate the objective function; hence, we only consider the protocol time to be the sum of phases 2–5.
Furthermore, our model excludes the processing latency of individual nodes, as this time is independent of the partitioning configuration and thus does not affect the optimization outcome.

\section{Committee Configuration Optimization (CCO)}\label{sec:cco-model}
In this section, we develop the CCO model to enhance consensus performance (i.e., minimize latency) by optimizing the committee configuration. 
We model the committee configuration optimization as an MIP model. 
In the following, we specify the objective function and the set of constraints. 


\subsection{Decision Variables and Objective Function}
Let $p$ be an integer decision variable representing the number of consensus committees, which is equivalent to the number of leaders. In the TopBFT protocol, each partition is required to maintain \(3f + 1\) nodes to ensure fault tolerance, consisting of \(2f + 1\) active nodes and \(f\) passive nodes. In the normal case, only the \(2f + 1\) active nodes participate in the consensus process within each committee. However, in the event of a TEE failure, the \(f\) passive nodes become active and also participate in the consensus, resulting in the classic \(3f + 1\) configuration in each committee.

Thus, our committee configuration decision variables are categorized into two types. The first type is the partition variable, which defines membership within a given committee. Let \( x_{ij} \) be a binary decision variable such that \( x_{ij} = 1 \) if \( i \) serves as the leader of \( j \), and \( x_{ij} = 0 \) otherwise, for all \( i, j \in \mathcal{S}_c, i \neq j \). For \( i = j \), \( x_{ii} = 1 \) indicates that \( i \) is designated as a leader of one of the formed committees, whereas \( x_{ii} = 0 \) implies that it functions as a follower. The second type is the connection variable, representing communication relationships during protocol operation. Let \( y_{ij} \) be a binary decision variable such that \( y_{ij} = 1 \) if the leader node \( i \) communicates with the follower node \( j \), and \( y_{ij} = 0 \) otherwise. We denote $d_{ij} > 0$ as the communication delay for the link $(i, j)$, while $d_{iv}$ and $d_{vi}$ represent the delays from node $i$ to the verification committee and vice versa.

We then formulate the latency of one transaction $T_{tr}$ as the objective function and minimize it. A successful transaction $T_{tr}$ in the consensus network can be represented as the sum of five time periods: 1) $T_{pre}$: time spent in \emph{pre-prepare} and \emph{prepare} phases; 2) $T_{cv}$: time to send the prepared message from the consensus committee to $S_v$; 3) $T_{ver}$: time to verify the message, reach an agreement in $S_v$ and form a block; 4) $T_{vc}$: time to communicate the valid message from $S_v$ to consensus committees; 5) $T_{com}$: time spent in the \emph{commit} phase in the consensus committee to validate the block from $S_v$.
This means that:
\begin{align}
    \min \ & \ T_{tr} \\
           & \ {T}_{tr} =  \ T_{pre} + \ T_{cv} + \ T_{ver} + \ T_{vc}+ T_{com}
\end{align}

\subsection{Constraints}
\subsubsection{Protocol Operation Latency Constraints}
$T_{pre}$ corresponds to the time spent in the two round trip times (RTT)s between the followers and their leader, which can be formulated as:
\begin{equation} \label{eq:pre-time_2}
    \forall {i,j} \in \mathcal{S}_c : \  T_{pre} \  \geq 2 \cdot \big( y_{ij} \cdot d_{ij} + y_{ij} \cdot d_{ji} \big)  
\end{equation}

$T_{cv}$ corresponds to the time spent on communicating messages from the consensus leader to the verification committee, thus we have: 
\begin{equation} \label{eq:cv-time_2}
    \forall {i} \in \mathcal{S}_c: \  T_{cv} \   \ \geq x_{ii} \cdot d_{iv}  
\end{equation}

In the verification process based on $S_v$, there is a total of four message RTTs, so we take four times the maximum value of the communication delay between the leader and follower members in the verification committee. 
\begin{equation} \label{eq:ver-time_2}
    \forall {i,j} \in \mathcal{S}_v :T_{ver}  \geq 4 \cdot \big( x_{ij} \cdot d_{ij} + x_{ij} \cdot d_{ji} \big)
\end{equation}

Similar to $T_{cv}$, $T_{vc}$ corresponds to the time spent on communicating messages from the verification committee to the consensus leader, thus we have: 
\begin{equation} \label{eq:vc-time_2}
    \forall {i} \in \mathcal{S}_c: \  T_{vc} \   \geq \   x_{ii} \cdot d_{vi}
\end{equation}

The time spent in the \emph{commit} phase requires two RTTs, then we have:
\begin{equation} \label{eq:com-time_2}
    \forall {i,j} \in \mathcal{S}_c : \  T_{com} \geq 2 \cdot \big( y_{ij} \cdot d_{ij} + y_{ij} \cdot d_{ji} \big)  
\end{equation}

\subsubsection{Node Set partition}
We now state the constraints related to the formation of parallel committees, based on the selection of leaders and followers. Firstly, each committee should have exactly one leader. Consequently, the number of committees corresponds to the number of leaders, implying the following constraint:
\begin{equation} \label{eq:milp-constr7_2}
    \sum_{i=1}^{N_c} x_{ii} = {p} 
\end{equation}

Secondly, for each committee, the followers only communicate with their respective leaders and do not communicate with the leaders from other committees. 
Thus we have the following constraints:
\begin{equation} \label{eq:milp-constr8_2} 
  \forall i,j \in \mathcal{S}_c: \  x_{ij} \leq x_{ii} 
\end{equation}
The left side of constraints~(\ref{eq:milp-constr8_2}) can take the value of 1 only when $x_{ii} = 1$. When $x_{ii} =0$, the node $i$ serves as a follower in a consensus committee; hence it cannot be the leader. Constraints~(\ref{eq:milp-constr8_2}) prevent any follower from communicating with nodes that are not their leader (leaders and followers of other committees).

Thirdly, any node belongs to one and only one committee (either as a leader or follower).  Consequently, a node $j$ can either be a follower of exactly one leader in its committee or the leader of this committee. Thus, we have:
\begin{equation} \label{eq:milp-constr9_2} 
    \forall j \in \mathcal{S}_c: \ \sum_{i=1}^{N_c} x_{ij} = 1 
\end{equation}

\subsubsection{Security Constraints}

The TopBFT network requires that each committee within the partitioned system contains at least \( 3f + 1 \) nodes, where \( f \) represents the minimum number of faults the committee must tolerate. This requirement ensures that at least \( \frac{2}{3} \) of the nodes in each committee are honest, thereby guaranteeing both safety and liveness in the consensus process.  Accordingly, the partitioning decision variable must meet the following constraints:
\begin{equation} \label{eq:milp-constr10_2}
    \forall i \in \mathcal{S}_c: \ \sum_{j=1}^{N_c} x_{ij} + 1 \geq (3 f + 1)x_{ii}
\end{equation}

We denote the Byzantine failure rate and crash failure rate of node $i$ as $b_i$ and $c_i$, respectively. 
Let $B$ and $C$ be the parameters representing the tolerated crash and Byzantine failure rates for leaders, respectively. 
This ensures that the leaders are selected from the most reliable and stable nodes: i.e.\ their failure rates $b_i$ and $c_i$ are lower than the system-tolerated thresholds $B$ and $C$, respectively.
We then formulate the corresponding constraints as follows:
\begin{align}
    \forall i \in \mathcal{S}_c: x_{ii} b_i \leq B \label{eq:milp-constr11_2} \\
    \forall i \in \mathcal{S}_c: x_{ii} c_i \leq C \label{eq:milp-constr12_2}
\end{align}

\subsubsection{Relationship between Partition Variable and Connection Variable}
In a configured committee, the connection between a leader node and a follower node must depend on both nodes being assigned to the same partition. This ensures that only nodes within the same partition can communicate directly within the consensus protocol. Thus, if two nodes are in different partitions, any direct connection between them is disallowed. This relationship is formalized by the following constraints, which enforces that connections are only allowed within the same committee:
\begin{equation} \label{eq:milp-contr13_2}
    \forall i,j \in \mathcal{S}_c: \  y_{ij} \leq x_{ij}
\end{equation} 
These constraints specifies that \( y_{ij} = 1 \) (indicating a connection) is permissible only if \( x_{ij} = 1 \), meaning both nodes \( i \) and \( j \) belong to the same partition.

\subsubsection{Adaptive Optimization for Fallback}
In normal case operations, TopBFT is designed to tolerate up to $f$ failures out of $2f+1$ nodes to ensure safety. Since TEEs ensure the monotonicity of the counter used for each sequence number, nodes cannot assign the same sequence number to different messages. Thus, based on the above security constraints, every proposed request can be guaranteed to correspond to the identical secret and consistent sequence number on all correct nodes, when the number of required participating nodes (i.e. lower bound) is reduced from $3f+1$ to $2f+1$ for each committee.

The worst-case scenario occurs when TEEs fail, which is equivalent to the presence of non-TEE nodes in the system. In this situation, the committees must run a classical BFT consensus protocol without TEEs, ensuring at least $2f+1$ correct nodes out of all $3f+1$ in a committee. Therefore, TopBFT includes a fallback process until the TEE failure is recovered.

To capture TEE status of each node and committee, we introduce a parameter \( t_i \) to represent the TEE condition of node \( i \). Specifically, \( t_i = 1 \) indicates that node \( i \) has experienced a TEE failure, while \( t_i = 0 \) indicates that node \( i \) is functioning normally without a TEE failure.  
Additionally, we define a variable \( \sigma_i \) to capture the overall TEE status of the committee led by node \( i \). Here, \( \sigma_i = 1 \) denotes that the committee headed by node \( i \) is impacted by a TEE failure, while \( \sigma_i = 0 \) indicates either that there is no TEE failure within the committee or that node \( i \) has not been chosen as the leader. 

Based on these, we formulate the following constraints:
\begin{align}
    & \forall i \in \mathcal{S}_c: \ t_ix_{ii}+t_jx_{ij} \leq N \sigma_i \label{eq:milp-constr14_2} \\
    & \forall i \in \mathcal{S}_c: \sum_{j=1}^{N_c} y_{ij} \geq (2+\sigma_i)f \label{eq:milp-constr15_2}
\end{align}
Constraints (\ref{eq:milp-constr14_2}) ensure that TEE failures are accurately represented across committee assignments, while constraints (\ref{eq:milp-constr15_2}) adjust the required number of follower nodes based on the TEE status of each committee, ensuring adequate fault tolerance in both normal and TEE failure scenarios.

\section{Experimental Evaluation}
We implemented the CCO model and integrate it with TopBFT testbed. The testbed is developed in C language, which builds an Intel SGX-enabled TEE for each node to implement the trusted monotonic counters. In the testbed design, each node can be the leader or follower running in the verification committee or one of the consensus committees and performs consensus operations by exchanging P2P messages under the TCP/IP protocol. We deployed the protocols on the same real-world cloud servers based on the Microsoft cloud platform, utilizing five Azure virtual machines (VMs), each VM with 8 vCPUs and 64 GB of RAM. Our consensus protocol and optimization model are executed by setting up at most 240 node instances distributed among the five VMs. The CCO model is implemented and solved by MIP optimization solver Cplex \cite{cplex}.

The key metrics to measure the protocol performance are based on: \emph{latency} (i.e., \emph{milliseconds, ms}) measured as the time spent from sending a group of client requests (i.e., transactions) to accepting valid replies for all requests in the group at the client, and \emph{throughput}, which is the number of operation messages processed every second in the system, denoted \emph{operations per second, op/s}. To implement the adversaries, we need to allow a subset of nodes (at most 30\% of all) to perform faulty operations, including long delays mimicking slow nodes and crash faults for nodes that then have no subsequent operations.

\subsection{Performance of Various BFT Algorithms}
\begin{figure} [!t]
     \centering
     \vspace{-3mm}
     \begin{subfigure}{0.45\textwidth}
         \centering
         \includegraphics[width=\textwidth]{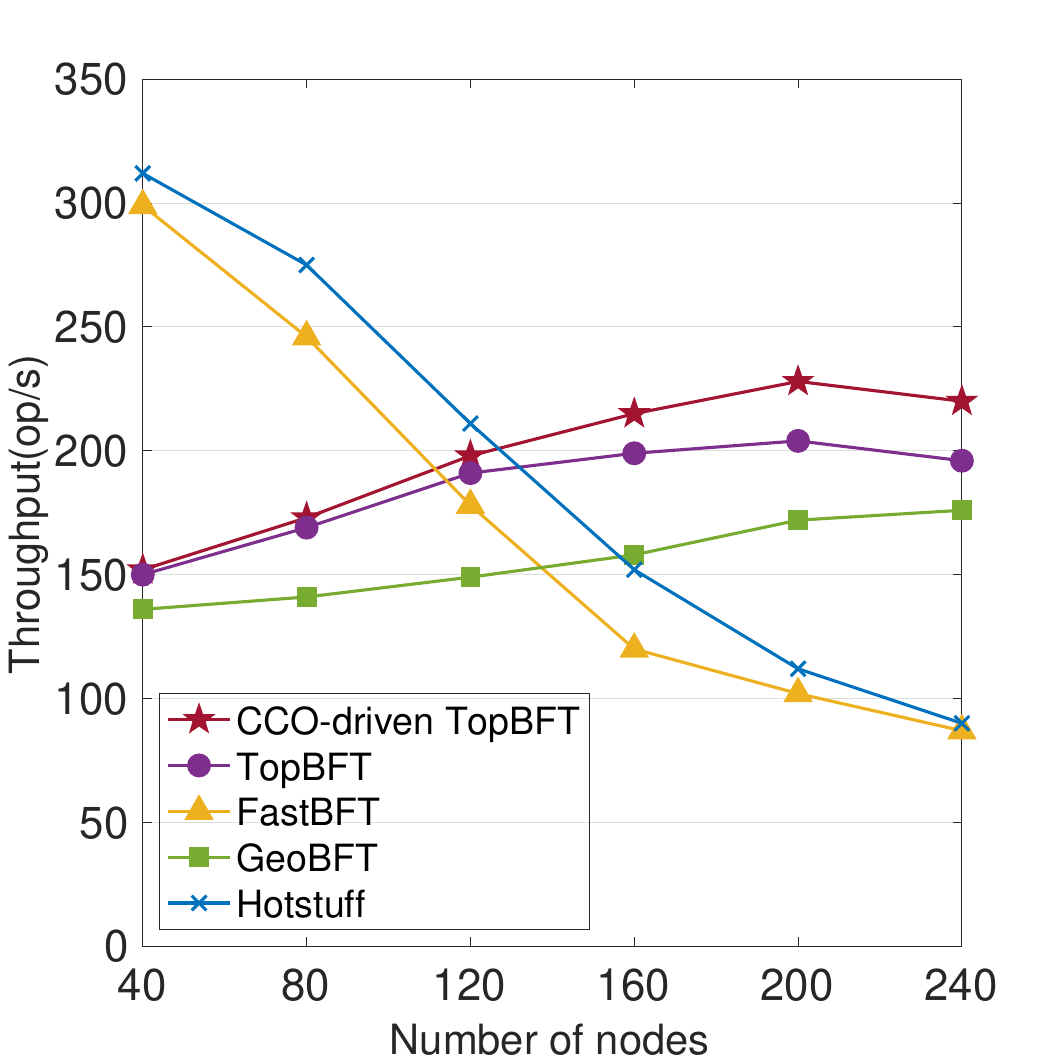}
         \caption{Throughput comparison}
         \label{ro-subfig1}
     \end{subfigure}
     \begin{subfigure}{0.45\textwidth}
         \centering
         \includegraphics[width=\textwidth]{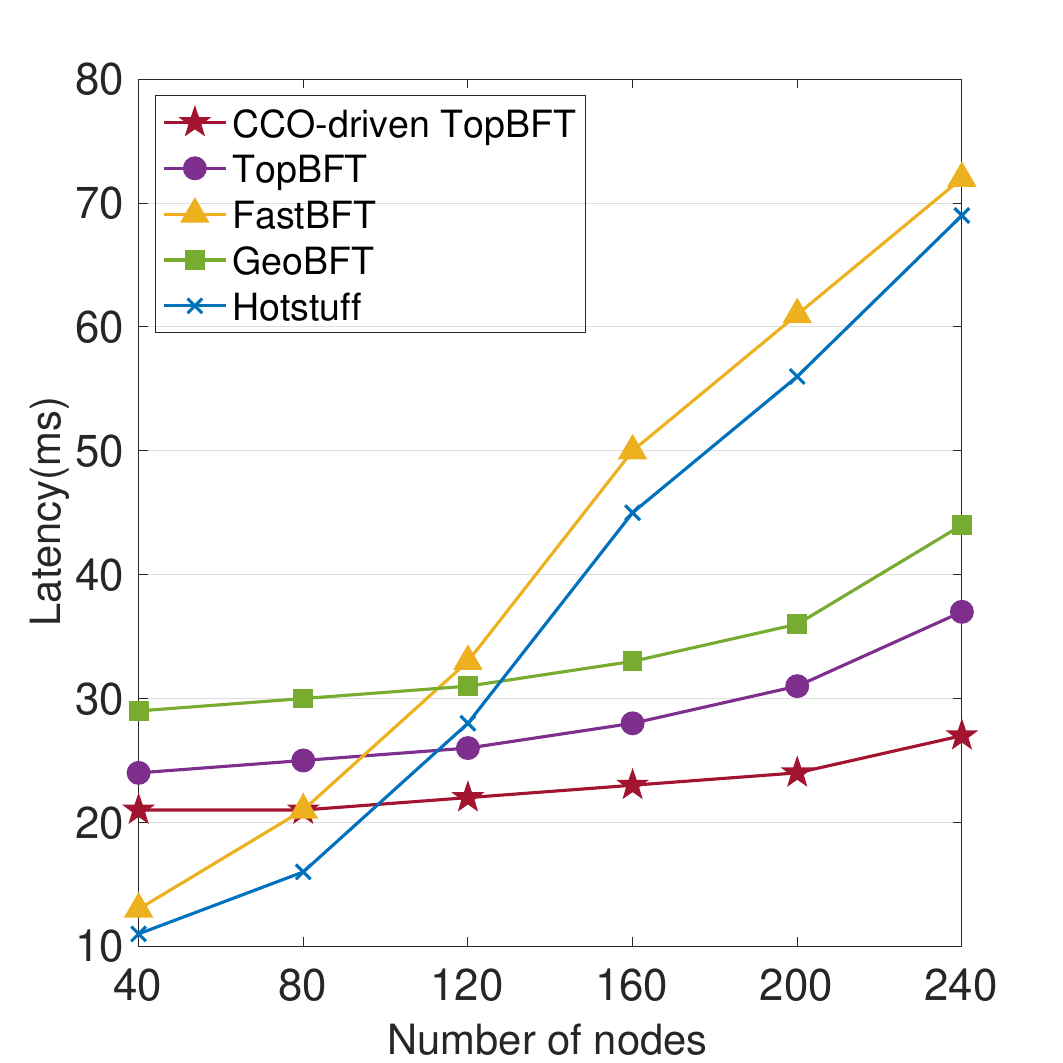}
         \caption{Latency comparison}
         \label{ro-subfig2}
     \end{subfigure}
        \caption{Performance comparison of BFT with various node size.}
        \label{ro-fig1}
        \vspace{-3mm}
\end{figure}

In this section, we evaluate the performance of normal case operations using a 1 MB operation size varying numbers of nodes from 40 to 240. We compare the performance between TopBFT and optimized TopBFT driven by CCO. The performance evaluation also compares with three other state-of-the-art BFT protocols chosen to give a broad comparison: HotStuff outperforms the traditional BFT-SMaRt \cite{yin2019hotstuff,bessani2014state}; FastBFT outperforms MinBFT and CheapBFT \cite{liu2018scalable}; and a parallel BFT scheme, i.e. GeoBFT \cite{gupta2020resilientdb}.
Fig. \ref{ro-subfig1} presents the average throughput of them. As the number of nodes increases, FastBFT and HotStuff both have a linear decline in consensus throughput down to around 80 op/s. This is because both protocols are designed to execute BFT consensus relying on a single committee in which all nodes must participate in the consensus operations to reach an agreement for each set of requests. Thus, when the number of nodes increases from 40 to 200, TopBFT outperforms FastBFT and HotStuff with throughput that rises to 200 op/s and remains stable thereafter. This throughput improvement is due to the consensus parallelism that allows each consensus committee to accept requests and run consensus independently and simultaneously --- when the number of nodes grows, more consensus committees will be created for these new nodes. Although GeoBFT also utilizes a parallel consensus scheme, its average throughput is still lower than (non-optimized) TopBFT due to inefficient consensus operations.

Here we have set up TopBFT to use four nodes randomly allocated to each committee, which is considered a ``default scheme'' for committee configuration compared to the CCO-driven configuration optimization. When the number of nodes goes up, the overall throughput goes up as well until the verification committee is fully utilized. This is shown by the steady throughput when the number of nodes goes from 200 to 240. In addition, when the model executes to optimize the committee configuration, each committee adopts the optimized organization scheme, including the optimized committee numbers and node allocation as designed in the CCO model. As a result, when applying the CCO model to a set of 240 nodes, the throughput can be improved up to around 230 op/s, which indicates a 15\% improvement compared to using the default scheme. 

Fig. \ref{ro-subfig2} shows the average consensus latency in the same condition as Fig. \ref{ro-subfig1}. With the increase in nodes from 40 to 240, FastBFT and HotStuff both suffer from linearly increasing consensus delays (up to approx 70 ms) since the growing number of nodes results in increased message exchanges between the leader and follower nodes despite their {\small$O(n)$} message complexity. However, when running TopBFT, each consensus committee independently accepts requests and executes consensus in parallel, so the delay can be measured based on each parallel consensus committee. Specifically, this delay should remain stable when each consensus committee has the same number of nodes (i.e., four nodes per committee by default) and the verification committee does not reach its full utilization.  When the verification committee is fully utilized (i.e., over 200 nodes), the consensus delay increases significantly under the default committee configuration. This is because more requests queue up in the verification committee and wait for verification and total ordering, resulting in longer consensus delays. However, when applying CCO to compute an optimized configuration scheme, the consensus delays of running TopBFT can be reduced to nearly a stable level, outperforming both non-optimized TopBFT and GeoBFT, as illustrated by the line in the graph representing the optimized TopBFT.

\subsection{Performance Comparison with Varying Operation Size}
\begin{figure} [!t]
     \centering
     \vspace{-3mm}
     \begin{subfigure}{0.45\textwidth}
         \centering
         \includegraphics[width=\textwidth]{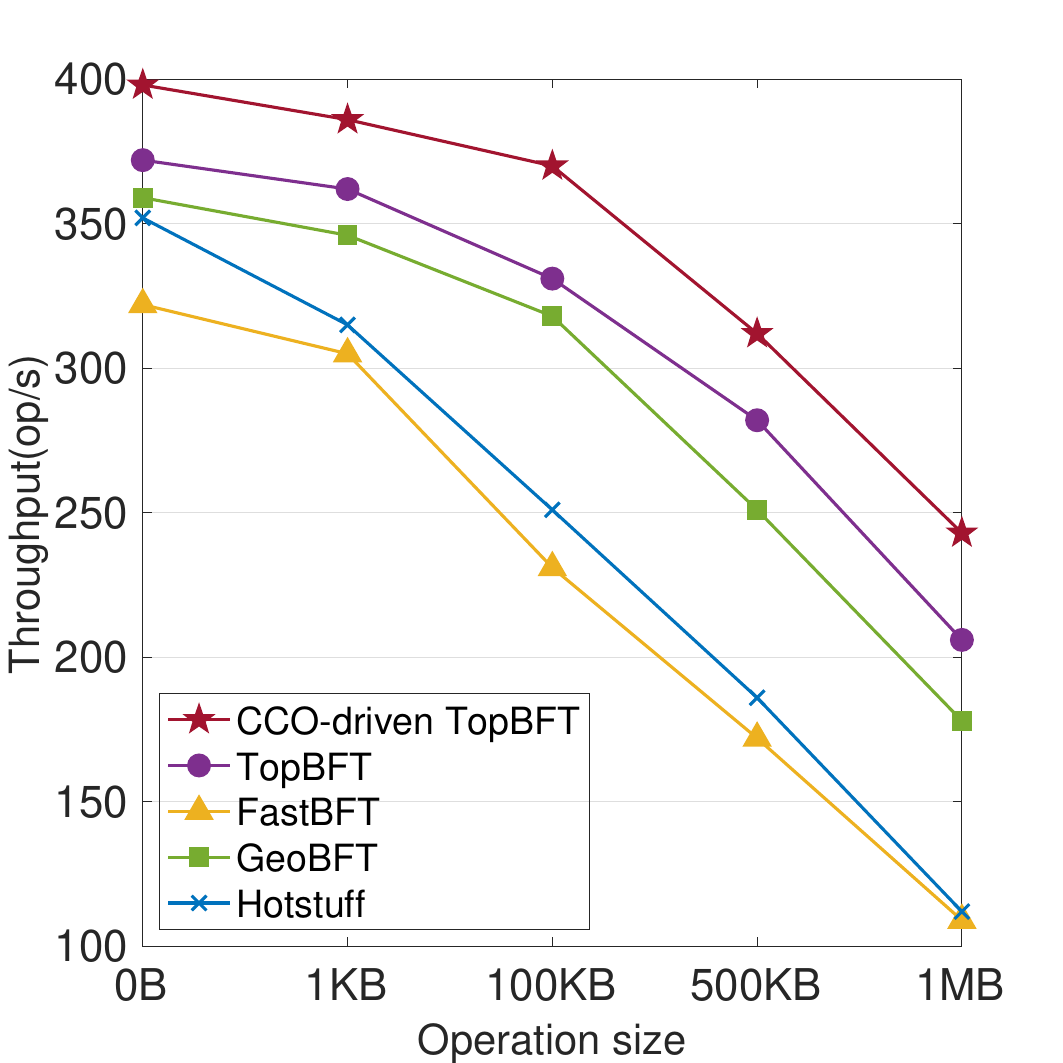}
         \caption{Throughput comparison}
         \label{ro-subfig3}
     \end{subfigure}
     \begin{subfigure}{0.45\textwidth}
         \centering
         \includegraphics[width=\textwidth]{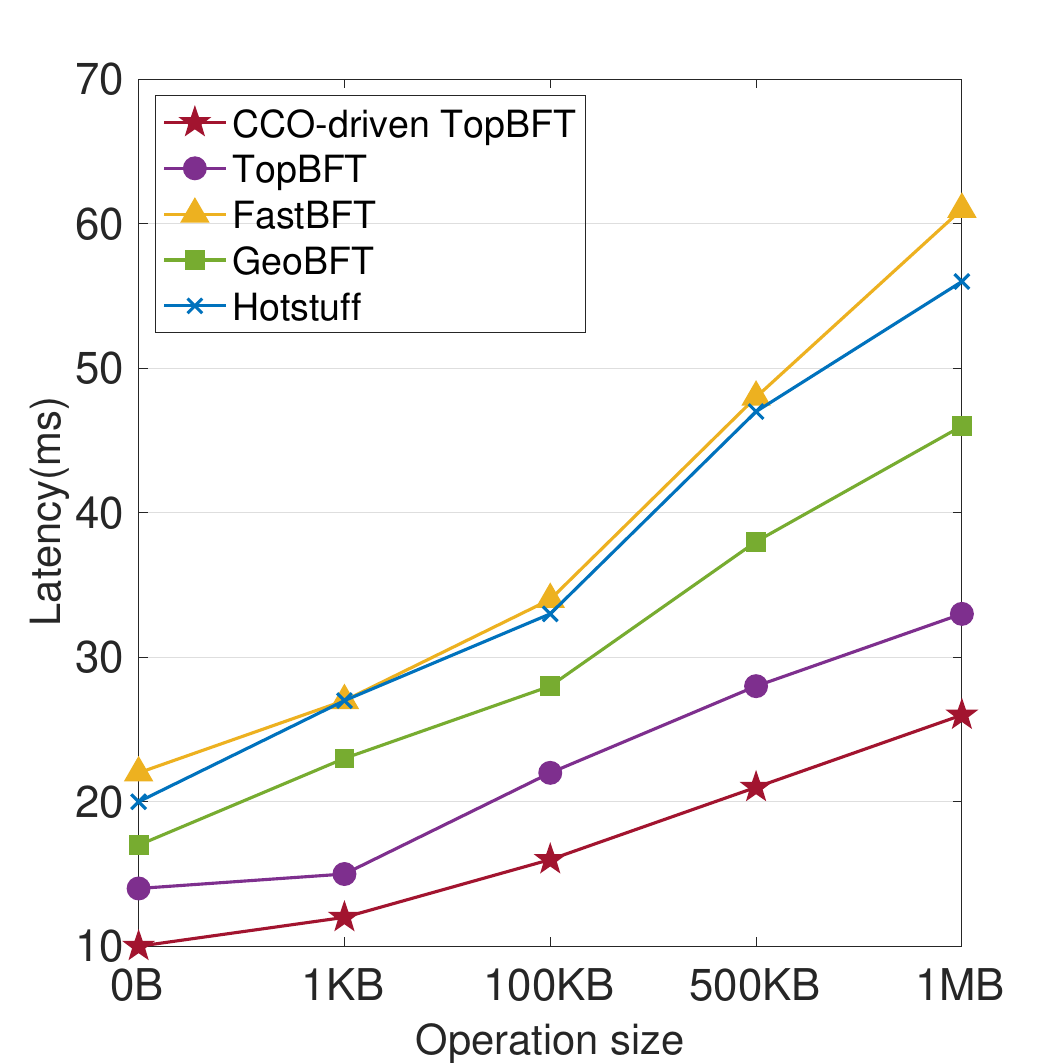}
         \caption{Latency comparison}
         \label{ro-subfig4}
     \end{subfigure}
        \caption{Performance comparison of BFT with various operation size.}
        \label{ro-fig2}
    \vspace{-3mm}
\end{figure}

Figs. \ref{ro-subfig3} and \ref{ro-subfig4} show the average throughput and latency of various consensus protocols, including CCO-driven TopBFT, TopBFT, GeoBFT, FastBFT, and Hotstuff, under operation payloads ranging from 0 to 1MB with 200 nodes. As the operation payload increases, all protocols exhibit a decline in throughput and an increase in latency, primarily due to the additional transmission delays caused by larger payload sizes. Despite this trend, the CCO-driven TopBFT consistently outperforms other protocols across both metrics, demonstrating significant performance improvements.

In Fig. \ref{ro-subfig3}, the throughput performance highlights the advantage of incorporating the CCO model into TopBFT. With smaller payloads, the difference between the protocols is smaller, as the message sizes impose minimal delays. Standard TopBFT maintains moderate throughput under larger payloads, the CCO-driven TopBFT achieves up to 29.4 \% higher throughput, particularly at payloads near 1MB. This improvement is attributed to the optimized committee configurations provided by the CCO model, which effectively reduces communication overhead and enhances processing efficiency.

Fig. \ref{ro-subfig4} presents the latency performance of the protocols as the operation payload grows. The CCO-driven TopBFT demonstrates lower latency compared to the other protocols. For instance, with a payload size of 1MB, the latency of CCO-driven TopBFT is approximately 18.9 \% lower than that of standard TopBFT and significantly outperforms GeoBFT, Hotstuff, and FastBFT. 

These results emphasize the effectiveness of the CCO model in enhancing the scalability and responsiveness of TopBFT, particularly in scenarios involving larger operation payloads. By optimizing the configuration of consensus committees taking failure rates and delays into account, the CCO-driven TopBFT achieves superior throughput and latency compared to existing protocols.

\subsection{Performance of Adaptive Fallback Optimization}
In this section, we present a performance comparison of TopBFT when operating under a fallback scenario. To implement the fallback, we set a fixed percentage of $30\%$ failed TEEs within the total node. In fallback scenarios, all $3f+1$ nodes in each committee participate in consensus operations to ensure fault tolerance. We compare a randomly allocated fallback, where nodes are assigned to committees without optimization, against the CCO-driven TopBFT fallback. In the CCO-driven TopBFT fallback, the configuration is adaptive: only committees containing failed TEEs transition to the $3f+1$ requirement, while healthy committees continue operating with $2f+1$ nodes.

\begin{figure} [!t]
     \centering
     \vspace{-3mm}
     \begin{subfigure}{0.45\textwidth}
         \centering
         \includegraphics[width=\textwidth]{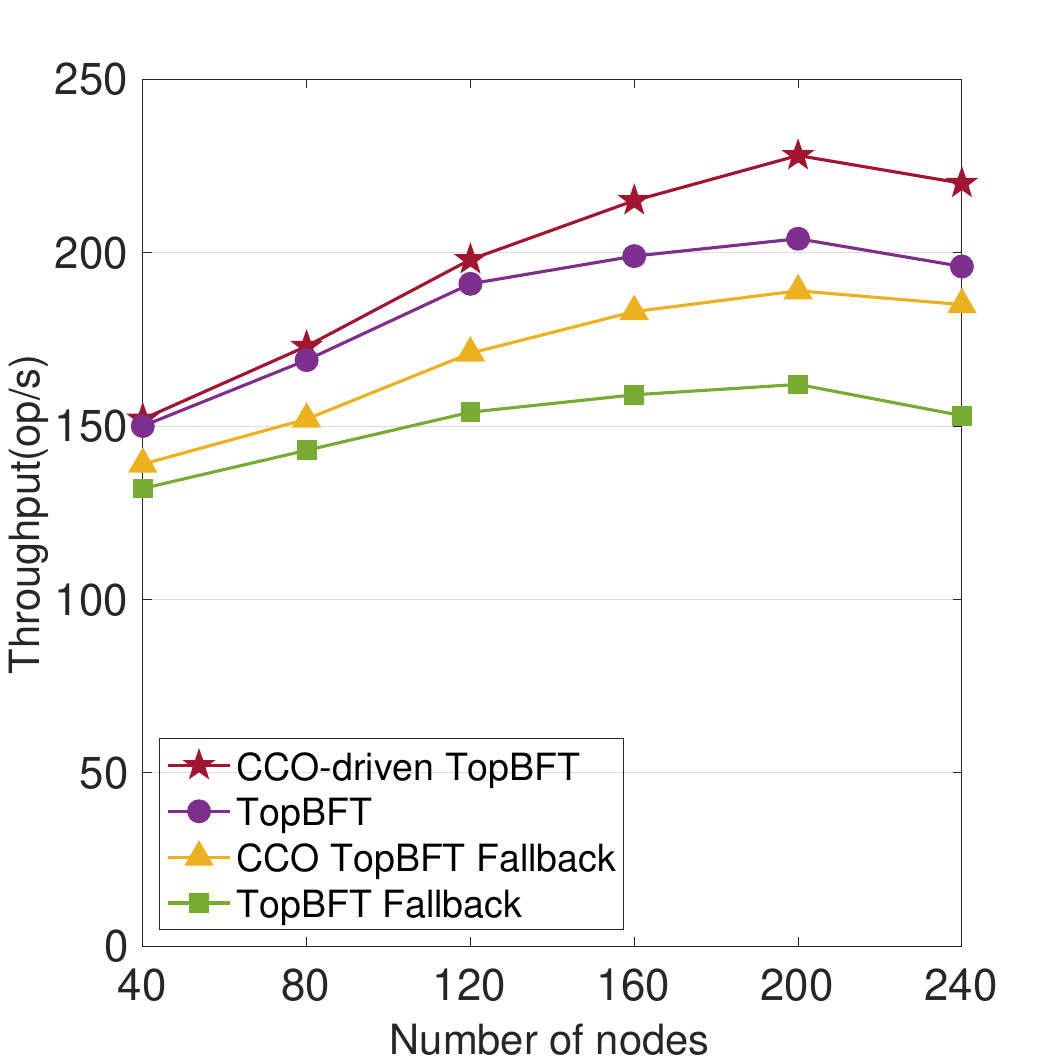}
         \caption{Throughput comparison}
         \label{ro-subfig5}
     \end{subfigure}
     \begin{subfigure}{0.45\textwidth}
         \centering
         \includegraphics[width=\textwidth]{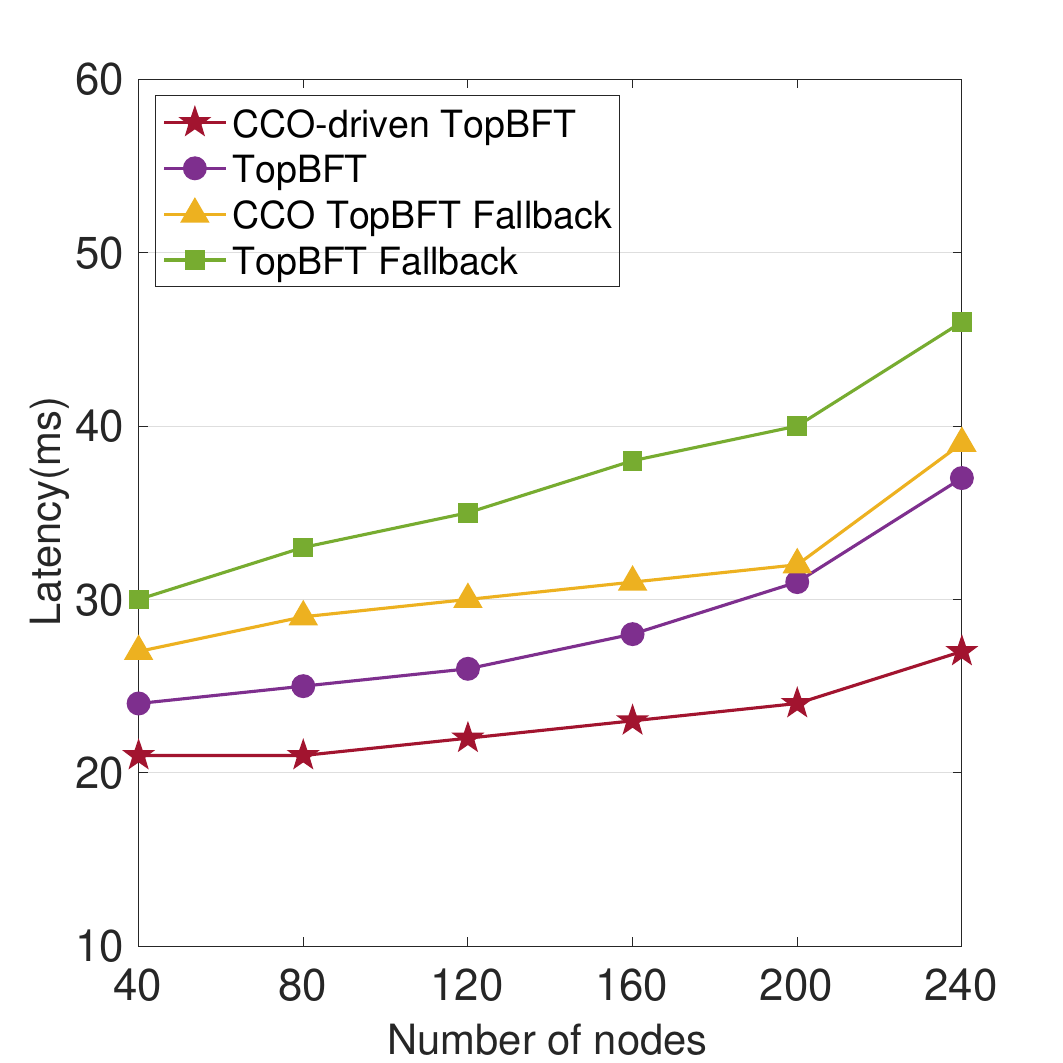}
         \caption{Latency comparison}
         \label{ro-subfig6}
     \end{subfigure}
        \caption{Performance comparison including fallback with various node size.}
        \label{ro-fig3}
    \vspace{-3mm}
\end{figure}

Fig. \ref{ro-subfig5} illustrates the throughput comparison. The results show that the CCO-driven TopBFT achieves approximately 21\% higher throughput compared to the standard fallback TopBFT. This improvement demonstrates the effectiveness of the CCO model in optimizing consensus operations under fallback conditions. However, both fallback configurations exhibit lower throughput compared to TopBFT under normal conditions, reflecting the additional overhead introduced by fallback operations.

Fig. \ref{ro-subfig6} presents the latency comparison. Both fallback mechanisms incur higher latency than the normal-case TopBFT due to the increased communication and reconfiguration requirements. Nevertheless, the CCO-driven fallback TopBFT shows reduced latency compared to the standard fallback approach, highlighting its ability to minimize delays and maintain more efficient operation under TEE failure scenarios.

These results highlight that the CCO-driven TopBFT fallback achieves superior performance compared to the standard fallback, offering improved performance and robustness in TEEs.

\section{Conclusion}

In this paper, we propose an CCO model for improving the performance of the trusted parallel BFT consensus algorithm. The model not only enhances fault tolerance and resilience but also optimally determines the connection configuration for each committee based on the current TEE status. This model ensures that the network can dynamically reconfigure its committee participant in consensus process during the fallback when TEE failures occur, maintaining both safety and liveness.
Experimental results demonstrate the significant effectiveness of our proposed model in TopBFT, showing notable performance improvements in both normal operations and scenarios involving TEE failures. The CCO-driven TopBFT consistently outperforms other configurations under diverse network conditions. These results underscore the model's capability to maintain secure and efficient parallel BFT consensus in TEEs. Consequently, our work provides a substantial enhancement to parallel BFT in TEEs, offering a promising solution for secure and high-performance blockchain applications.

\begin{credits}

\subsubsection{\discintname}
The authors have no competing interests to declare that are relevant to the content of this paper.
\end{credits}
%
%
%
\bibliographystyle{splncs04}
\bibliography{mainref}

\end{document}